\documentclass[reprint]{JASA}

\usepackage{bbding}
\usepackage{booktabs} 
\usepackage{multirow}

\begin{document}
 

\title[Retrospective PAQs and continuous P/E]{Understanding the relationship between retrospective soundscape attributes and continuous pleasantness and eventfulness ratings}

\author{Yuanbo Hou}
\email{Yuanbo.Hou@eng.ox.ac.uk}
\affiliation{Department of Engineering Science, University of Oxford, Oxford OX1 3PJ, United Kingdom}

\author{Kening Guo}
\affiliation{UCL Institute for Environmental Design and Engineering, The Bartlett, University College London, London WC1H 0NN, United Kingdom}

\author{Francesco Aletta}
\affiliation{UCL Institute for Environmental Design and Engineering, The Bartlett, University College London, London WC1H 0NN, United Kingdom}

\author{Woon-Seng Gan}
\affiliation{School of Electrical and Electronic Engineering, Nanyang Technological University, Singapore 639798, Singapore}

\author{Dick Botteldooren}
\affiliation{Department of Information Technology, Faculty of Engineering and Architecture, Ghent University, 9052 Ghent, Belgium}

\author{Jian Kang}
\affiliation{UCL Institute for Environmental Design and Engineering, The Bartlett, University College London, London WC1H 0NN, United Kingdom} 



\begin{abstract} 
Soundscape studies ask listeners to rate perceived affective quality (PAQ) attributes, such as pleasant and vibrant, after listening to recordings or experiencing acoustic environments. These retrospective ratings summarise an experience during which impressions may change. International Organization for Standardization (ISO) procedures convert these PAQ ratings into Pleasantness and Eventfulness scores. Pleasantness and Eventfulness (P/E) can also be rated directly and continuously during listening, capturing changes in reported perception over time. Retrospective PAQ ratings and continuous direct P/E describe the same soundscape experience from different perspectives.
We examine how PAQs relate to direct P/E and how ISO-derived P/E corresponds to direct judgements, using 1,782 paired 30-s trials involving audiovisual recordings from 18 streets. 
The results show that PAQs retain a predominantly two-dimensional structure. Vibrant and Monotonous lie closer to the Eventfulness--Uneventfulness axis in the direct P/E plane than in the ISO 2025 reference geometry. 
The full 30-s direct P/E mean has lower calibrated error against ISO-derived P/E than three 10-s means. 
An empirical mapping using the same 8 PAQs reduces held-out joint root-mean-square error by 4.9\% relative to affine-calibrated ISO 2025 scoring. The findings show that retrospective and continuous assessments offer related but distinct soundscape descriptions.
\end{abstract} 

\maketitle

\section{Introduction}\label{sec:introduction}  
Beyond sound level measurements, the soundscape approach considers how people perceive, experience, and understand acoustic environments in context \cite{ISO12913_1_2014}. Soundscape assessment complements physical acoustic measurements by examining the human experience of sound. This experience is multifaceted, shaped by acoustic characteristics, temporal dynamics, individual responses, and environmental context \cite{brown2011standardisation,kang2016tenquestions,aletta2016descriptors}. How listeners are asked to assess an acoustic environment can affect whether their responses reflect an overall impression or changes in perception during listening.

Soundscape assessment commonly uses perceptual ratings to characterise how people experience an acoustic environment, including ratings of affective attributes organised around Pleasantness and Eventfulness \cite{Axelsson2010,ISO12913_2_2018,mitchell2022,hou2023jasa}.
When listeners evaluate an exposure as a whole after it has ended, their responses constitute a retrospective overall assessment: a summary judgement of an experience extending over a period of time. This approach is relatively straightforward to administer and can be applied across different environments and exposure durations, but does not directly capture how perception changes over time. In contrast, continuous assessment records listeners' perceptions as an experience unfolds, allowing temporal changes to be examined. However, it requires sustained attention and repeated responses, which may increase participant burden and become difficult to maintain over longer exposures. The two approaches offer complementary perspectives, with different practical advantages and limitations in capturing overall and time-varying soundscape perception.

In practice, retrospective overall and continuous soundscape assessments can be implemented through different approaches to measuring Pleasantness and Eventfulness.

The ISO framework derives Pleasantness and Eventfulness (P/E) from ratings of 8 perceived affective quality (PAQ) attributes: pleasant, vibrant, eventful, chaotic, annoying, monotonous, uneventful, and calm. Under ISO/TS 12913-3:2019, these ratings are projected onto two orthogonal dimensions using trigonometric weights based on attribute directions at 45° intervals \cite{ISO12913_3_2019}. ISO/TS 12913-3:2025 uses language-specific adjusted attribute directions, including for English \cite{ISO12913_3_2025}. The framework builds on Russell's circumplex model of affect \cite{russell1980circumplex,russell1981affective} and its application to soundscape perception \cite{Axelsson2010}. Subsequent studies have identified a quasi-circumplex structure, with some deviations from the assumed angular positions \cite{aletta2024soundscape,blattermann2026bayesian}. When used retrospectively, ISO P/E provides an overall appraisal but is less practical for continuous tracking.

In contrast, direct P/E allows listeners to report their perceived Pleasantness and Eventfulness directly within a two-dimensional space, without deriving coordinates from separate attribute ratings. Building on dimensional affect measurement, including the Affect Grid \cite{russell1989affectgrid}, continuous response methods enable perceptual changes to be tracked as an experience unfolds \cite{schubert1999continuously,nagel2007emujoy}. In soundscape research, such approaches have been used to examine temporal variations in pleasantness \cite{aumond2017continuous}, compare continuous and retrospective assessments \cite{graetzer2020continuous}, and investigate responses to changing acoustic characteristics and sound sources \cite{han2022foreground, han2023sequence}. 

If retrospective overall assessments can reflect aspects of how perception develops over time, understanding their relationship with continuous assessments could have important implications for soundscape research and design. In laboratory experiments and soundwalks, researchers must decide how long participants should listen before providing an overall assessment. For example, ISO/TS 12913-2:2018 gives three minutes as an example of a listening period during soundwalks, yet it remains unclear how listening duration relates to the overall judgement formed. Does an overall assessment primarily reflect the average perception throughout an exposure, or do particular moments and temporal changes also shape the final judgement? In soundscape design, understanding when a perceptual response emerges and how long it persists can complement the overall evaluation of an intervention.

The extent to which an overall judgement reflects these temporal aspects of perception, however, may depend on the characteristics of the soundscape and the conditions of assessment. During three-minute simulated urban walks, mean continuous pleasantness principally explained overall pleasantness \cite{aumond2017continuous}. Recency and trend effects provided additional explanatory value for controlled sequences, but not for the natural audiovisual sequences examined. For 30-second soundscapes, continuous pleasantness--eventfulness ratings and post-exposure semantic differential assessments revealed broadly similar differences between soundscape categories \cite{graetzer2020continuous}. Principal component analysis of the post-exposure ratings identified dimensions corresponding to pleasantness and eventfulness. At a longer timescale, retrospective soundscape pleasantness was associated with average momentary evaluations, peak experiences, temporal trends, and mood at recall \cite{Steffens2017}. These findings support treating an overall judgement as an appraisal formed across an experience, while examining empirically how it relates to the temporal course of reported perception.

\begin{figure}[t]
\centering
\includegraphics[width=0.5\textwidth]{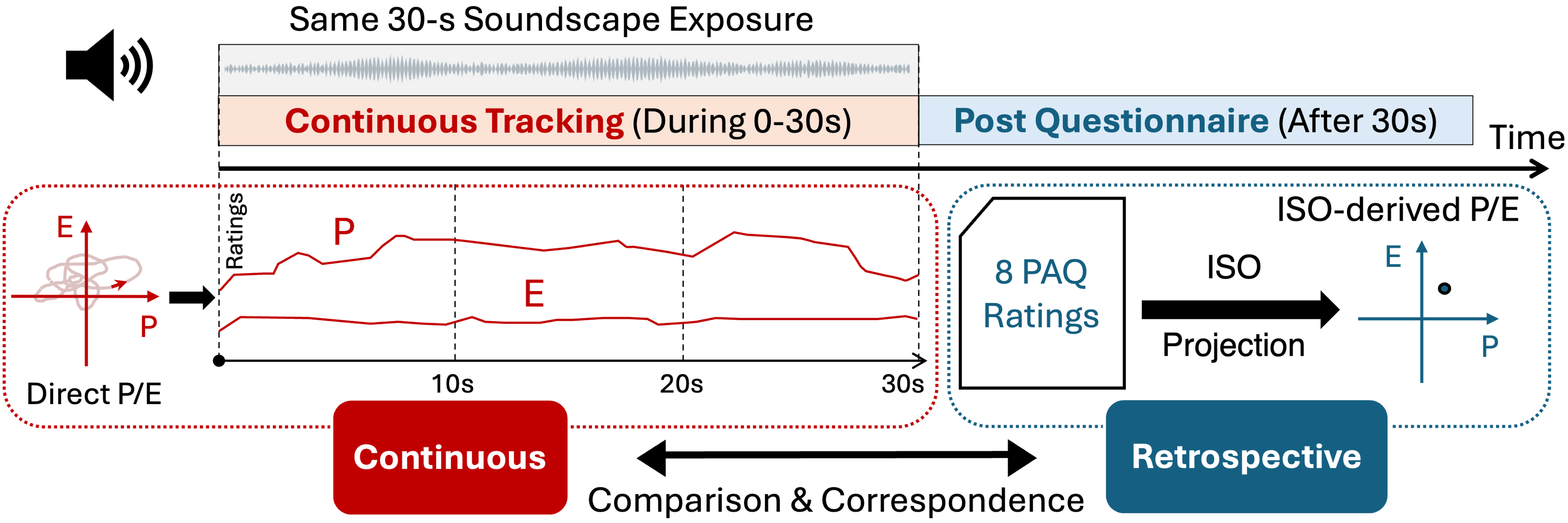}
\caption{Paired retrospective and continuous assessments of the same 30-s soundscape exposure. During listening, participants report Pleasantness and Eventfulness (direct P/E) continuously by moving a cursor within a two-dimensional P/E display, producing time-resolved ratings. After listening, participants rate the 8 perceived affective quality (PAQ) attributes, which are converted into ISO-derived P/E coordinates. The continuous records are summarised over time for trial-level comparison with the retrospective scores.}
\label{fig:paired_soundscape_assessments}
\end{figure}

While retrospective overall and continuous assessments have shown some correspondence, questions remain about how their perceptual representations and numerical ratings relate, and whether a more flexible PAQ-to-P/E mapping could better estimate average direct P/E. To examine these relationships, this study uses paired retrospective PAQ ratings and continuous P/E ratings from the same listeners for the same soundscape exposures \cite{guo2026movement}. Fig.~\ref{fig:paired_soundscape_assessments} illustrates the two assessment approaches and their relationship. During each 30-second exposure, participants continuously reported their perceived Pleasantness and Eventfulness, followed immediately by retrospective ratings of the 8 PAQ attributes. These paired observations allow the relationships between the two assessment approaches to be examined in terms of perceptual structure, numerical correspondence, and empirical mapping, leading to three research questions:\\
\textit{RQ1: How do the affective qualities captured by the 8 retrospective PAQ ratings relate to one another and to Pleasantness and Eventfulness reported continuously during listening?}\\
\textit{RQ2: Which period or summary of continuous ratings is closest to the retrospective overall assessment, and how much do they differ?}\\
\textit{RQ3: Can a more flexible scoring rule use retrospective PAQ ratings to better estimate the average assessment during listening?}

The remainder of this paper is organised as follows. Section~\ref{sec:Method} describes the paired measurements and analysis methods. Section~\ref{sec:result_analysis} first characterises the data (Section~\ref{sec:section_A}). Sections~\ref{sec:section_B}--\ref{sectionD} examine the relationships among retrospective PAQs and their associations with direct P/E to answer RQ1. Section~\ref{sectionE} examines temporal correspondence and trial-level agreement to answer RQ2. Section~\ref{sectionF} evaluates empirical PAQ-to-P/E mapping to answer RQ3. Section~\ref{sec:discussion} discusses these findings in relation to the three research questions, and Section~\ref{sec:Conclusion} concludes the paper.

\section{Method}\label{sec:Method} 

\subsection{Dataset and paired perceptual measurements}\label{sec:method_data}

The study involved 34 participants who assessed audiovisual stimuli representing 18 urban streets under three visual conditions: a static image (S), a standing-view video (SV), and a walking-view video (WV) \cite{guo2026movement}. The same street-specific audio recording was used across the three visual conditions. Each participant experienced all 54 street--condition combinations in a randomised order.
Before the main trials, an explanatory video and guided practice trials introduced the PAQ attributes and the response interface \cite{guo2026movement}.
Among the 34 participants, one participant's data were excluded as invalid, leaving 33 participants with usable paired records.
The present analysis uses these 33 participants across 18 streets and three visual conditions, yielding $33 \times 18 \times 3 = 1{,}782$ paired trials with retrospective PAQ ratings and continuous direct P/E records.
Table~\ref{tab:data_configuration} reports the study configuration and the completeness of the recorded direct P/E trajectories.

Each trial lasted 30~s. During the exposure, participants reported Pleasantness and Eventfulness (direct P/E) by moving a cursor within a two-dimensional P/E display. 
Cursor positions were recorded at 8~Hz, yielding up to 240 P/E coordinate pairs per trial.
Immediately afterwards, participants rated the overall soundscape on a 1--5 scale using the 8 perceived affective quality (PAQ) attributes: pleasant, vibrant, eventful, chaotic, annoying, monotonous, uneventful, and calm. 
The direct P/E records and retrospective PAQ ratings correspond to the same participant, street, and visual condition.

\begin{table}[t]
\caption{Configuration of the paired retrospective PAQ and continuous direct P/E measurements.}
\label{tab:data_configuration}
\centering 
\setlength{\tabcolsep}{6pt}
\renewcommand{\arraystretch}{0.6} 
\begin{tabular}{@{}lr@{\hspace{8pt}}lr@{}}
\toprule
\multicolumn{2}{c}{\textbf{Retrospective PAQ}} &
\multicolumn{2}{c}{\textbf{Direct P/E}} \\
\cmidrule(lr){1-2}\cmidrule(lr){3-4}
PAQ attributes       & 8     & P/E dimensions                & 2     \\
Rating scale         & 1--5  & Duration / trial              & 30~s  \\
Paired PAQ sets      & 1,782 & Sampling rate                 & 8~Hz  \\
Sets / condition     & 594   & Samples / trial               & 240   \\
                     &       & Complete trajectories         & 1,724 \\
                     &       & $\geq$80\% valid trajectories & 1,777 \\
\bottomrule
\end{tabular}
\end{table}

For comparisons that require one direct P/E coordinate per trial, the continuous P and E records are averaged separately across the 30-s exposure. Only samples with valid values for both coordinates enter either mean:
\begin{equation}
\setlength{\abovedisplayskip}{3pt}
\setlength{\belowdisplayskip}{3pt} 
\bar{P}^{\mathrm{direct}}_k=\frac{1}{n_k}\sum_{t\in\mathcal{V}_k}P_{k,t},
\qquad
\bar{E}^{\mathrm{direct}}_k=\frac{1}{n_k}\sum_{t\in\mathcal{V}_k}E_{k,t}
\label{eq:direct_mean30}
\end{equation}
where $k=1,\ldots,1782$ indexes trials; $P_{k,t}$ and $E_{k,t}$ are the P and E coordinates at sample $t$; $\mathcal{V}_k$ contains the time points with valid P and E values; and $n_k=|\mathcal{V}_k|$. The 30-s means provide one coordinate pair per trial; the 8-Hz samples are not treated as independent trials. The full continuous records are retained for comparisons of different temporal summaries.

The retrospective route uses the same 8 PAQ ratings to calculate P/E using the ISO 2019 and ISO 2025 reference geometries. Let $q_{k,i}$ be the 1--5 response to PAQ $i$ in trial $k$. Under ISO/TS~12913-3:2019, the 8 attribute directions are separated by $45^\circ$ \cite{ISO12913_3_2019}, giving
\begin{equation}
\setlength{\abovedisplayskip}{3pt}
\setlength{\belowdisplayskip}{3pt}
\resizebox{0.9\linewidth}{!}{$
\begin{aligned}
P^{2019}_k &= \frac{1}{4+\sqrt{32}}
\left[(P-A)+\frac{\sqrt{2}}{2}(C_a-C_h)+\frac{\sqrt{2}}{2}(V-M)\right]\\
E^{2019}_k &= \frac{1}{4+\sqrt{32}}
\left[(E-U)+\frac{\sqrt{2}}{2}(C_h-C_a)+\frac{\sqrt{2}}{2}(V-M)\right]
\end{aligned}
$}
\label{eq:iso2019}
\end{equation}
where $\mathrm{P}$, $\mathrm{V}$, $\mathrm{E}$, $\mathrm{C_h}$, $\mathrm{A}$, $\mathrm{M}$, $\mathrm{U}$, and $\mathrm{C_a}$ denote the ratings of Pleasant, Vibrant, Eventful, Chaotic, Annoying, Monotonous, Uneventful, and Calm, respectively, for trial $k$. ISO/TS~12913-3:2025 instead prescribes language-specific adjusted attribute directions. The adjusted-English angles, in the same PAQ order, are $0^\circ$, $46^\circ$, $94^\circ$, $138^\circ$, $177^\circ$, $231^\circ$, $275^\circ$, and $340^\circ$ \cite{ISO12913_3_2025,aletta2024soundscape}. After centring each rating at the response-scale midpoint, the ISO 2025 coordinates are
\begin{align}
\setlength{\abovedisplayskip}{3pt}
\setlength{\belowdisplayskip}{3pt} 
P^{2025}_k=\frac{1}{\lambda_P}\sum_{i=1}^{8}x_{k,i}\cos\theta_i, \quad \lambda_P=\frac{\rho}{2}\sum_{i=1}^{8}|\cos\theta_i|. \\
E^{2025}_k=\frac{1}{\lambda_E}\sum_{i=1}^{8}x_{k,i}\sin\theta_i, \quad \lambda_E=\frac{\rho}{2}\sum_{i=1}^{8}|\sin\theta_i|.
\label{eq:iso2025}
\end{align}
where $x_{k,i}$ denotes the PAQ rating after midpoint centring, $\theta_i$ its adjusted-English direction, and $\rho=4$ for the 1--5 scale. These ISO 2025 coordinates correspond to $P_{\mathrm{ISO}}$ and $E_{\mathrm{ISO}}$ in ISO/TS 12913-3:2025; the year superscripts distinguish the two ISO scoring rules examined in this study. Each trial provides the direct P/E coordinate pair $(\bar{P}^{\mathrm{direct}}_k,\bar{E}^{\mathrm{direct}}_k)$ and two ISO-derived P/E pairs calculated from the same 8 retrospective PAQ ratings under the ISO 2019 and ISO 2025 scoring rules.

\subsection{Internal and cross-method geometric analysis}\label{sec:method_geometry}

The internal geometry of the 8 retrospective PAQs is examined independently of direct P/E, with the ISO 2019 and 2025 angles serving as references. 
For each participant, the 54 ratings of each PAQ are centred on that participant's mean for the attribute and standardised by the corresponding pooled standard deviation.
The resulting 1,782 trial-level observations give an $8\times8$ Pearson correlation matrix $\mathbf{C}$. Principal component analysis (PCA) identifies the two dominant dimensions through the eigenvalues $\lambda_m$ and eigenvectors $\mathbf{v}_m$:
\begin{equation}
\setlength{\abovedisplayskip}{3pt}
\setlength{\belowdisplayskip}{3pt}
\mathbf{C}\mathbf{v}_m=\lambda_m\mathbf{v}_m 
\label{eq:paq_pca}
\end{equation}
where $\mathbf{v}_m=(v_{\mathrm{P},m},\allowbreak v_{\mathrm{V},m},\allowbreak v_{\mathrm{E},m},\allowbreak v_{\mathrm{Ch},m},\allowbreak v_{\mathrm{A},m},\allowbreak v_{\mathrm{M},m},\allowbreak v_{\mathrm{U},m},\allowbreak v_{\mathrm{Ca},m})^{T}$, and $m=1,\ldots,8$.
The first two principal components give each PAQ two loading coordinates. Multiplying each eigenvector coefficient by the square root of its eigenvalue gives
\begin{equation}
\setlength{\abovedisplayskip}{3pt}
\setlength{\belowdisplayskip}{3pt}
\mathbf{l}_i=(l_{i,1},l_{i,2})=\left(
v_{i,1}\sqrt{\lambda_1},
v_{i,2}\sqrt{\lambda_2}
\right), \quad
i=1,\ldots,8 
\label{eq:paq_loading}
\end{equation}

The 8 loading vectors form $\mathbf{L}\in\mathbb{R}^{8\times2}$:
\begin{equation*}
\setlength{\abovedisplayskip}{3pt}
\setlength{\belowdisplayskip}{3pt}
\resizebox{1\linewidth}{!}{$
\displaystyle
\mathbf{L}=
\begin{bmatrix}
v_{\mathrm{P},1}\sqrt{\lambda_1} &
v_{\mathrm{V},1}\sqrt{\lambda_1} &
v_{\mathrm{E},1}\sqrt{\lambda_1} &
v_{\mathrm{Ch},1}\sqrt{\lambda_1} &
v_{\mathrm{A},1}\sqrt{\lambda_1} &
v_{\mathrm{M},1}\sqrt{\lambda_1} &
v_{\mathrm{U},1}\sqrt{\lambda_1} &
v_{\mathrm{Ca},1}\sqrt{\lambda_1}
\\
v_{\mathrm{P},2}\sqrt{\lambda_2} &
v_{\mathrm{V},2}\sqrt{\lambda_2} &
v_{\mathrm{E},2}\sqrt{\lambda_2} &
v_{\mathrm{Ch},2}\sqrt{\lambda_2} &
v_{\mathrm{A},2}\sqrt{\lambda_2} &
v_{\mathrm{M},2}\sqrt{\lambda_2} &
v_{\mathrm{U},2}\sqrt{\lambda_2} &
v_{\mathrm{Ca},2}\sqrt{\lambda_2}
\end{bmatrix}^{\top}
$}
\label{eq:paq_loading_matrix_transpose}
\end{equation*}

To compare directions, each row of $\mathbf{L}$ is normalised to unit length, giving $\mathbf{U}\in\mathbb{R}^{8\times2}$.
PCA does not fix the orientation of the resulting plane relative to the P/E axes. An orthogonal alignment fits the whole empirical configuration to each ISO geometry, using a single rotation or reflection for all 8 directions:
\begin{equation}
\setlength{\abovedisplayskip}{3pt}
\setlength{\belowdisplayskip}{3pt}
\widetilde{\mathbf{U}}=\mathbf{U}\mathbf{T}
\label{eq:paq_alignment}
\end{equation}
where $\mathbf{T}\in\mathbb{R}^{2\times2}$ minimises the distance between $\mathbf{U}\mathbf{T}$ and the prescribed ISO unit directions. The transformation preserves the relative PAQ angles. After separate alignments to ISO 2019 and ISO 2025, the shortest circular difference for each attribute describes its deviation from the corresponding reference direction.

The PAQ correlations alone do not establish how the directly rated P and E dimensions are oriented relative to either ISO reference geometry.
For this comparison, both the 30-s direct P/E means and the retrospective PAQs are participant-centred and standardised by their pooled standard deviations. Let $\mathbf{X}\in\mathbb{R}^{1782\times2}$ contain the direct P/E pairs and $\mathbf{Q}\in\mathbb{R}^{1782\times8}$ contain the matched PAQ ratings. The joint regression is
\begin{equation}
\setlength{\abovedisplayskip}{3pt}
\setlength{\belowdisplayskip}{3pt}
\mathbf{Q}=\mathbf{X}\mathbf{B}^{\mathsf T}+\mathbf{R}
\label{eq:direct_axis_regression}
\end{equation}
where each row of $\mathbf{B}\in\mathbb{R}^{8\times2}$ contains the coefficients for regressing one PAQ rating on direct P and E jointly, and $\mathbf{R}$ contains the regression residuals. A dimension aligned with a given direction in the circumplex has a cosine-shaped association with attributes around the circle. For PAQ item $i$, let $\theta_i^{\mathrm{ISO}}\in[0^\circ,360^\circ)$ be its prescribed reference angle \cite{Axelsson2010,ISO12913_3_2025} and $\mathbf{u}_i^{\mathrm{ISO}}=(\cos\theta_i^{\mathrm{ISO}},\sin\theta_i^{\mathrm{ISO}})^{\mathsf T}$ its unit-direction vector. For each direct P/E dimension $D\in\{\mathrm{P},\mathrm{E}\}$, the 8 regression coefficients are fitted at these reference directions using a cosine-profile summary \cite{GurtmanBalakrishnan1998,ZimmermannWright2017}:
\begin{equation}
\setlength{\abovedisplayskip}{3pt}
\setlength{\belowdisplayskip}{3pt}
\resizebox{0.9\linewidth}{!}{$
\displaystyle
\beta_{D,i}\approx a_D+\mathbf{u}_i^{\mathrm{ISO}\mathsf T}\mathbf{v}_D
=a_D+b_D\cos\theta_i^{\mathrm{ISO}}+c_D\sin\theta_i^{\mathrm{ISO}}
$}
\label{eq:direct_axis_cosine}
\end{equation}
where $i=1,\ldots,8$, $a_D$ is the offset, and $b_D$ and $c_D$ define the directional cosine component $\mathbf{v}_D=(b_D,c_D)^{\mathsf T}$. The least-squares fit uses a fixed $360^\circ$ cosine period and weights the 8 prescribed PAQ angles equally. The fitted direction is $\widehat{\phi}_D=\operatorname{atan2}(c_D,b_D)$, measured counterclockwise from the positive P axis. The cosine-profile coefficients $a_D$, $b_D$, and $c_D$ are fitted separately using the ISO 2019 and ISO 2025 reference angles, while the joint regression matrix $\mathbf{B}$ remains unchanged. Shortest signed angular differences from $0^\circ$ for P and $90^\circ$ for E locate the direct axes relative to each reference. Confidence intervals are obtained from 2,000 participant-level bootstrap resamples, with each participant's repeated trials resampled together.

The fitted axes summarise the association of each direct P/E dimension with the whole PAQ set. Individual PAQ directions require the reverse geometric reference: the axes now remain fixed to direct P/E. The two-predictor regression is fitted again with the PAQs participant-centred and standardised as above, but with direct P/E participant-centred in its original common P/E units, without separate P and E standardisation. Separate scaling would change the angle inferred from the two regression coefficients. For PAQ $i$, the re-estimated coefficients define its direct-P/E-anchored direction:
\begin{equation}
\setlength{\abovedisplayskip}{3pt}
\setlength{\belowdisplayskip}{3pt}
\widehat{\theta}_i^{\mathrm{direct}}=\operatorname{atan2}\left(\beta_{\mathrm{E},i},\beta_{\mathrm{P},i}\right),\qquad i=1,\ldots,8
\label{eq:direct_anchored_paq_angle}
\end{equation}

The shortest circular differences between these directions and the prescribed ISO 2019 or ISO 2025 angles describe the item-level deviations. The ISO angles serve as comparison references and are not used to estimate the regression coefficients in this direct-anchored analysis.

\subsection{Temporal summaries and trial-level agreement}\label{sec:method_agreement}

Retrospective PAQ ratings provide an overall assessment after listening, while continuous P/E ratings record judgements throughout the 30-s exposure.
For the temporal comparison, four summaries of the continuous record are examined: the full 30-s mean and the first, middle, and last 10-s means. The four reported summaries are evaluated against retrospective ISO 2025-derived P/E under leave-one-participant-out (LOPO) validation, using joint standardised root mean squared error (RMSE). Of the 1,782 paired trials, two were excluded from the temporal analysis: one had no valid observations during the first 10 s, and the other had none during the final 1 s. This leaves 1,780 trials in the common subset defined across the broader set of candidate temporal summaries, including the 1-s end-of-exposure summaries. Both excluded trials had valid full-duration means. For each summary, the P and E differences are divided by the standard deviation of the corresponding direct P/E coordinate before they are combined into the two-dimensional (2D) error.
The full 30-s direct P/E means, $\bar{P}^{\mathrm{direct}}_k$ and $\bar{E}^{\mathrm{direct}}_k$, defined in Eq.~(\ref{eq:direct_mean30}), are used for the main trial-level agreement and mapping comparisons.

The direct P/E interface and ISO projection produce coordinates on different numerical scales. An affine calibration estimates a separate offset and slope for the P and E axes of ISO 2025-derived P/E:
\begin{equation}
\setlength{\abovedisplayskip}{3pt}
\setlength{\belowdisplayskip}{3pt}
D_k^{\mathrm{ISO2025-cal}}=a_D+b_D D_k^{\mathrm{ISO2025}},\qquad D\in\{\mathrm{P},\mathrm{E}\}
\label{eq:iso2025_affine_calibration}
\end{equation}
where $D_k^{\mathrm{ISO2025}}$ is the ISO 2025-derived P or E coordinate for trial $k$, $D_k^{\mathrm{ISO2025-cal}}$ is its calibrated value on the direct P/E scale, and $a_D$ and $b_D$ are the calibration parameters estimated only from training trials. In LOPO validation, the trials from one participant form the held-out fold; in leave-one-street-out (LOSO) validation, all trials from one street are held out. Each fitted calibration is shared across the three visual conditions within the fold. For the temporal comparison, separate LOPO calibrations are fitted against each temporal summary on its own scale.

Trial-level agreement is evaluated between calibrated ISO 2025-derived P/E coordinates and the full 30-s direct P/E means.
Lin's concordance correlation coefficient (CCC) measures correspondence with the line of equality, where the two measurements assign the same value \cite{Lin1989}. P and E RMSE describe their coordinate-wise discrepancies, while joint 2D RMSE combines the two errors in the original direct P/E coordinate units. 
Difference-versus-mean plots display the difference between calibrated ISO 2025-derived and direct P/E against the mean of the two measurements for each coordinate \cite{BlandAltman1986}. Within the P/E plane, quadrant agreement records whether P and E have the same signs under both measurements; the shortest absolute angle between the two coordinate vectors describes directional disagreement. The main agreement analysis uses all 1,782 trials with valid full-duration direct P/E means.

\subsection{Empirical PAQ-to-P/E mapping and held-out evaluation}\label{sec:method_relaxed}

The paired retrospective PAQ ratings and 30-s direct P/E means allow an empirical PAQ-to-P/E mapping to be estimated. 
The affine-calibrated ISO 2025 projection serves as the reference scoring rule for held-out comparison: calibration adjusts each output coordinate's scale and offset without changing the relative PAQ weights within that coordinate. 
The empirical mapping uses all 8 original PAQ ratings without first reducing them to differences between opposing attributes. 
For trial $k$, $\mathbf{y}_k\in[1,5]^8$ contains the retrospective PAQs, and $\mathbf{z}_k\in\mathbb{R}^2$ denotes a modelled P/E state on the scale of direct P/E. Their participant-centred relationship is represented by a linear-Gaussian measurement model: 
\begin{equation}
\setlength{\abovedisplayskip}{3pt}
\setlength{\belowdisplayskip}{3pt}
\mathbf{y}_k^{\mathrm{pc}}=\mathbf{W}\mathbf{z}_k^{\mathrm{pc}}+\boldsymbol{\epsilon}_k,\quad k=1,\ldots,1782
\label{eq:relaxed_measurement_model}
\end{equation}
where $\mathbf{W}\in\mathbb{R}^{8\times2}$ relates the P/E dimensions to each PAQ, and $\boldsymbol{\epsilon}_k$ represents correlated PAQ variation not described by the P/E state. The model allows Gaussian variation in the state and in the PAQ residuals. The observed 30-s direct P/E mean $\mathbf{d}_k$ is represented as $\mathbf{d}_k=\mathbf{z}_k+\boldsymbol{\eta}_k$, where $\boldsymbol{\eta}_k$ is direct-measurement variation. Direct P/E is used to estimate the measurement relationship but is not treated as an error-free state.

Equation~(\ref{eq:relaxed_measurement_model}) describes PAQ ratings conditional on a P/E state. Scoring a new PAQ response requires the inverse relationship. Under the Gaussian model, its conditional mean estimates P/E from the 8 retrospective PAQ ratings:
\begin{equation}
\setlength{\abovedisplayskip}{3pt}
\setlength{\belowdisplayskip}{3pt}
\widehat{\mathbf{z}}_k=\boldsymbol{\mu}_z+\boldsymbol{\Sigma}_z\mathbf{W}^{\mathsf T}\left(\mathbf{W}\boldsymbol{\Sigma}_z\mathbf{W}^{\mathsf T}+\boldsymbol{\Omega}\right)^{-1}\left(\mathbf{y}_k-\boldsymbol{\mu}_y\right)
\label{eq:relaxed_inverse_mapping}
\end{equation}
where $\boldsymbol{\mu}_z$ and $\boldsymbol{\Sigma}_z$ denote the mean and covariance of the modelled P/E state, $\boldsymbol{\mu}_y$ is the mean PAQ response, and $\boldsymbol{\Omega}$ combines residual PAQ covariance with between-participant response variation. All parameters and training-fold means are estimated from training data. The inverse mapping takes uncentred held-out PAQ ratings as input; their paired direct P/E values are reserved for evaluation and do not enter the prediction.

The empirical mapping is evaluated against the affine-calibrated ISO 2025 projection using identical LOPO and LOSO folds. For every held-out participant or street, RMSE is computed for P, E, and the joint 2D coordinates against direct P/E. The fold-specific errors are averaged, assigning equal weight to each held-out participant or street.

\section{\label{sec:result_analysis}Results and Analysis}  

The paired sample and measurement distributions are presented in Section~\ref{sec:section_A}. Sections~\ref{sec:section_B}--\ref{sectionD} examine the relationships among the 8 retrospective PAQs and their associations with direct P/E to answer RQ1. Section~\ref{sectionE} compares temporal summaries of continuous direct P/E with retrospective ISO-derived P/E and examines their trial-level agreement to answer RQ2. Section~\ref{sectionF} evaluates whether an empirical PAQ-to-P/E mapping improves the estimation of mean direct P/E relative to affine-calibrated ISO 2025 scoring to answer RQ3.

\subsection{Analysis sample and measurement distributions}\label{sec:section_A}

\begin{table}[t]
\caption{Distributions of retrospective PAQ ratings, direct P/E, and ISO-derived P/E. Values are mean (SD); $n=1{,}782$ overall and $n=594$ for each visual condition (S, SV, WV).}
\label{tab:measurement_distributions}
\centering
\setlength{\tabcolsep}{1pt}
\renewcommand{\arraystretch}{1}
\resizebox{0.5\columnwidth}{!}{%
\begin{tabular}{@{}lcccc@{}}
\toprule
\textbf{Measure} & \textbf{Overall} & \textbf{S} & \textbf{SV} & \textbf{WV} \\
\midrule
\multicolumn{5}{@{}l}{\textit{Retrospective PAQ ratings (1--5)}} \\
\hline
Pleasant    & 2.79(0.92) & 2.73(0.92) & 2.79(0.90) & 2.85(0.93) \\
Vibrant     & 3.13(0.84) & 3.11(0.84) & 3.13(0.86) & 3.15(0.83) \\
Eventful    & 3.26(0.89) & 3.26(0.87) & 3.25(0.90) & 3.27(0.91) \\
Chaotic     & 3.10(0.97) & 3.15(0.97) & 3.05(0.97) & 3.10(0.96) \\
Annoying    & 3.25(0.97) & 3.29(0.96) & 3.22(0.99) & 3.23(0.97) \\
Monotonous  & 2.89(0.92) & 2.89(0.92) & 2.90(0.93) & 2.89(0.92) \\
Uneventful  & 2.77(0.96) & 2.77(0.97) & 2.78(0.97) & 2.75(0.96) \\
Calm        & 2.55(0.99) & 2.51(0.97) & 2.57(0.99) & 2.57(1.00) \\
\addlinespace[1pt]
\toprule
\multicolumn{5}{@{}l}{\textit{Direct Pleasantness (P) / Eventfulness (E) (30-s mean)}} \\
\hline
P & $-0.185(0.441)$ & $-0.218(0.435)$ & $-0.180(0.442)$ & $-0.157(0.445)$ \\
E & 0.147(0.360) & 0.135(0.367) & 0.153(0.356) & 0.152(0.355) \\
\addlinespace[1pt]
\toprule
\multicolumn{5}{@{}l}{\textit{ISO 2019-derived P/E}} \\
\hline
P & $-0.069(0.287)$ & $-0.088(0.285)$ & $-0.062(0.284)$ & $-0.058(0.290)$ \\
E & 0.108(0.356) & 0.113(0.350) & 0.100(0.361) & 0.112(0.358) \\
\addlinespace[1pt]
\toprule
\multicolumn{5}{@{}l}{\textit{ISO 2025-derived P/E}} \\
\hline
P & $-0.080(0.293)$ & $-0.099(0.291)$ & $-0.072(0.290)$ & $-0.069(0.297)$ \\
E & 0.099(0.356) & 0.102(0.350) & 0.091(0.361) & 0.103(0.356) \\
\bottomrule
\end{tabular}
}
\end{table}

The experiment includes 34 participants and 18 London street environments under three visual conditions: static image (S), standing view (SV), and walking view (WV).
Among the 34 participants, one participant's data were excluded as invalid, resulting in 33 participants with 1,782 matched trials for the cross-method analysis, as summarised in Table~\ref{tab:data_configuration}.

Table~\ref{tab:measurement_distributions} summarises the trial-level distributions. The 8 retrospective PAQs have mean ratings ranging from 2.55 for Calm to 3.26 for Eventful, with similar means across S, SV, and WV. The direct 30-s means are $-0.185$ (SD = $0.441$) for P and $0.147$ (SD = $0.360$) for E. Both ISO projections have negative mean P and positive mean E coordinates and yield similar overall distributions. The distinction from direct P/E is clearest in Pleasantness: its standard deviation is $0.441$ for direct P/E, compared with $0.287$ under ISO 2019 and $0.293$ under ISO 2025, whereas the corresponding Eventfulness standard deviations are close ($0.360$ versus $0.356$ under either ISO projection). The two ISO scoring rules produce similar marginal distributions, but their Pleasantness coordinates show substantially less variation than direct P/E. 


\subsection{Internal structure of the 8 PAQs}\label{sec:section_B}  

The internal relationships among the 8 retrospective PAQs are examined independently of direct P/E to determine whether they support a predominantly 2D configuration. Their estimated directions are then compared with the ISO 2019 and ISO 2025 reference geometries to describe the arrangement of the PAQs in this dataset.

PCA of the 8-PAQ correlation matrix yields eigenvalues of $4.453$ and $1.660$ for the first two components. As the eigenvalues of this matrix sum to 8, these two components together account for $(4.453+1.660)/8=76.4\%$ of the total variance. The empirical PAQ directions derived from these components are separately aligned with the ISO 2019 and ISO 2025 reference geometries using the procedure in Section~\ref{sec:method_geometry}. 

\begin{figure}[t]
\centerline{\includegraphics[width=0.5\textwidth]{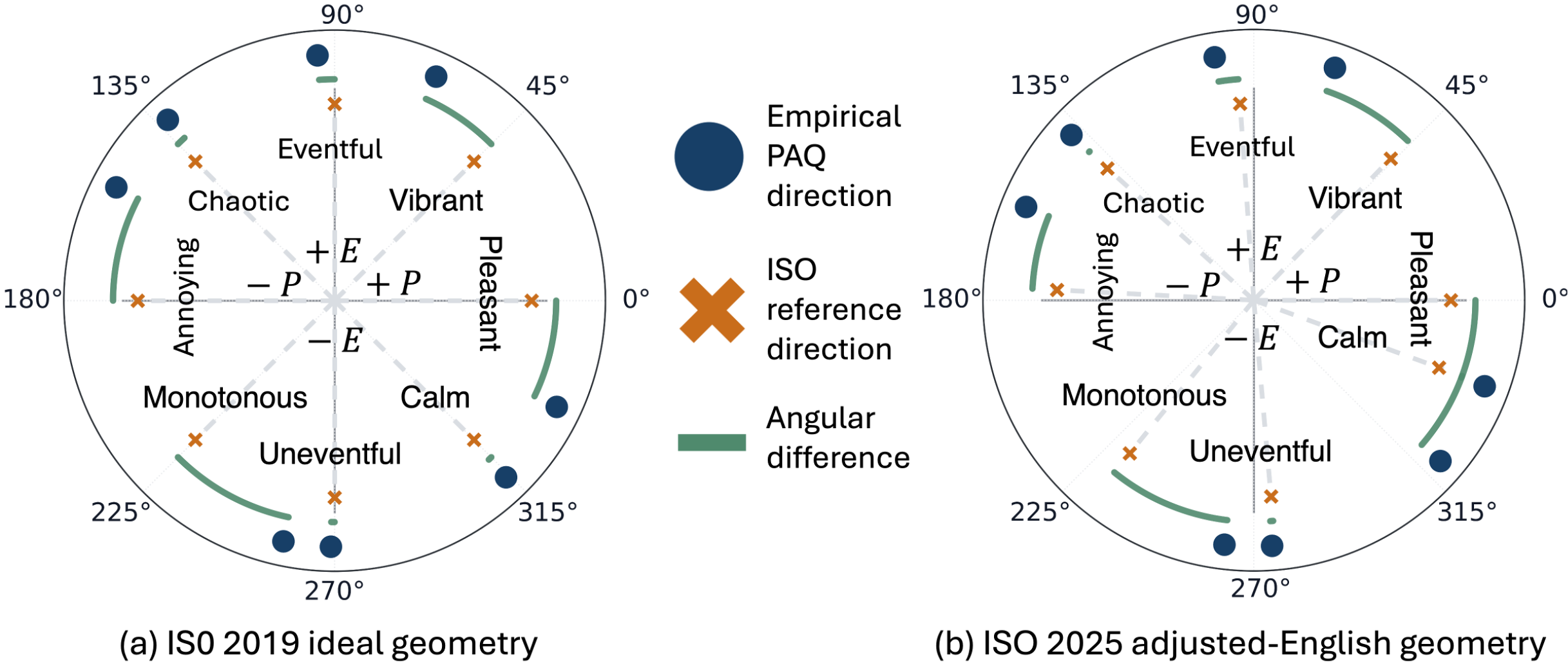}}
\caption{Internal 2D geometry of the 8 PAQs. Blue circles indicate the empirical PAQ directions, orange crosses indicate the corresponding ISO reference directions, and green arcs indicate the shortest angular differences between them. Radial offsets are used only to separate the markers visually; the comparison concerns angular direction. Grey diameters indicate the ISO Pleasantness and ISO Eventfulness axes.}
\label{fig:Fig2_Internal_PAQ_compact}
\end{figure}

Fig.~\ref{fig:Fig2_Internal_PAQ_compact} compares the empirical PAQ directions with the two ISO references after alignment. For PAQ item $i$, the empirical angle $\widehat{\theta}_i$ is calculated from the corresponding row of $\widetilde{\mathbf{U}}$ in Eq.~(\ref{eq:paq_alignment}); its angular deviation is the shortest signed circular difference from the corresponding ISO reference angle $\theta_i^{\mathrm{ISO}}$.

Table~\ref{tab:internal_paq_angles} reports the estimated PAQ directions relative to the ISO 2019 reference. Eventful, Chaotic, Uneventful, and Calm have small absolute deviations of $4.0^\circ$, $2.3^\circ$, $0.9^\circ$, and $0.9^\circ$, respectively, whereas Pleasant, Vibrant, Annoying, and Monotonous deviate by $25.6^\circ$, $20.7^\circ$, $27.4^\circ$, and $33.0^\circ$. The relative directions nevertheless retain much of the expected circumplex organisation: Pleasant--Annoying, Eventful--Uneventful, and Chaotic--Calm are separated by $178.1^\circ$, $175.1^\circ$, and $178.6^\circ$, respectively, close to the expected $180^\circ$. Vibrant--Monotonous has the largest departure from opposition, with a separation of $167.7^\circ$. The empirical PAQ configuration preserves most opposing-pair relationships even where individual directions differ substantially from their ISO reference angles.

\begin{table}[t]
\caption{Empirical PAQ directions compared with the ISO 2019 and ISO 2025 reference geometries (degrees). Ref./Est. gives the ISO reference angle and the PCA-derived estimate after alignment. $\Delta$ is the shortest signed difference (estimate minus reference); positive values indicate counterclockwise displacement.}
\label{tab:internal_paq_angles}
\centering 
\setlength{\tabcolsep}{3pt}
\renewcommand{\arraystretch}{0.9}
\begin{tabular}{@{}cccc@{\hspace{8pt}}ccc@{}}
\toprule
 & \multicolumn{3}{c}{\textbf{ISO 2019}} & \multicolumn{3}{c}{\textbf{ISO 2025}} \\
\cmidrule(lr){2-4}\cmidrule(l){5-7}
\textbf{PAQ} & \textbf{Ref.} & \textbf{Est.} & \textbf{$\Delta$} & \textbf{Ref.} & \textbf{Est.} & \textbf{$\Delta$} \\
\midrule
Pleasant    & $0$ & $334.4$    & $-25.6$ & $0$ & $339.6$    & $-20.4$ \\
Vibrant     & $45$ & $65.7$    & $+20.7$ & $46$ & $70.8$    & $+24.8$ \\
Eventful    & $90$ & $94.0$    & $+4.0$  & $94$ & $99.1$    & $+5.1$  \\
Chaotic     & $135$ & $132.7$  & $-2.3$  & $138$ & $137.8$  & $-0.2$  \\
Annoying    & $180$ & $152.6$  & $-27.4$ & $177$ & $157.7$  & $-19.3$ \\
Monotonous  & $225$ & $258.0$  & $+33.0$ & $231$ & $263.1$  & $+32.1$ \\
Uneventful  & $270$ & $269.1$  & $-0.9$  & $275$ & $274.3$  & $-0.7$  \\
Calm        & $315$ & $314.1$  & $-0.9$  & $340$ & $319.2$  & $-20.8$ \\
\bottomrule
\end{tabular}
\end{table}

Table~\ref{tab:internal_paq_angles} also reports the empirical PAQ directions relative to the ISO 2025 adjusted-English reference. Pleasant and Annoying have smaller absolute deviations than under ISO 2019, decreasing from $25.6^\circ$ to $20.4^\circ$ and from $27.4^\circ$ to $19.3^\circ$, respectively; Chaotic decreases from $2.3^\circ$ to $0.2^\circ$. The deviation of Vibrant increases from $20.7^\circ$ to $24.8^\circ$, while Monotonous has the largest deviation under ISO 2025 at $32.1^\circ$. Calm shows the largest change, with its absolute deviation increasing from $0.9^\circ$ under ISO 2019 to $20.8^\circ$ under ISO 2025. The same empirical PAQ configuration has smaller angular deviations for some attributes and larger deviations for others, depending on the reference geometry.

Table~\ref{tab:internal_geometry_fit} summarises the mean, median, and maximum absolute angular deviations of the empirical PAQ configuration relative to the two reference geometries. The mean and median deviations are lower relative to ISO 2019, while the maximum deviations are similar; individual attributes exhibit different patterns. 
The PAQs retain a predominantly 2D configuration with largely preserved opposite-pair relationships. 

\begin{table}[t]
\caption{Overall angular fit of the empirical PAQ geometry to the ISO 2019 and ISO 2025 reference geometries.}
\label{tab:internal_geometry_fit}
\centering
\footnotesize
\setlength{\tabcolsep}{1pt}
\renewcommand{\arraystretch}{0.6}
\begin{tabular}{c|c|c|c}
\toprule
\textbf{Reference} &
\textbf{Mean $|\Delta|$ ($^\circ$)} &
\textbf{Median $|\Delta|$ ($^\circ$)} &
\textbf{Maximum $|\Delta|$ ($^\circ$)} \\
\midrule
ISO 2019 & 14.35 & 12.33 & 32.99 \\
ISO 2025 & 15.44 & 19.87 & 32.13 \\
\bottomrule
\end{tabular}
\end{table}

\subsection{Alignment of direct P/E with the PAQ circumplex}\label{sectionC}

Fig.~\ref{fig:Fig2_Internal_PAQ_compact} and Table~\ref{tab:internal_geometry_fit} describe the internal directions of the 8 retrospective PAQs in this dataset. The paired direct P/E measurements allow a separate examination of how the directly reported P and E dimensions relate to these attributes within the reference circumplex.
 
The associations between the 8 PAQs and the directly reported P and E dimensions are estimated from the 1,782 paired trials. Cosine-profile fits then locate the direct P/E axes relative to the ISO 2019 and ISO 2025 reference geometries. The underlying PAQ--direct P/E regression coefficients remain unchanged across the two references; only the reference angles used to estimate the fitted directions differ. Angular deviations are measured from the nominal Pleasantness ($0^\circ$) and Eventfulness ($90^\circ$) axes, with confidence intervals obtained from participant-level bootstrap resampling.

\begin{table}[t]
\caption{Fitted directions of direct P/E relative to the ISO 2019 and ISO 2025 PAQ reference geometries. Fitted angles are measured counterclockwise from the positive P axis; signed $\Delta$ denotes the shortest angular difference from the nominal P ($0^\circ$) or E ($90^\circ$) axis. The 95\% confidence intervals (CIs) of $\Delta$ are estimated from 2,000 participant-level bootstrap resamples.}
\label{tab:direct_axis_alignment}
\centering
\footnotesize
\setlength{\tabcolsep}{1pt}
\renewcommand{\arraystretch}{0.6}
\begin{tabular}{c|c|c|c|c}
\toprule
\textbf{ISO} & \textbf{Direct P/E} & \textbf{Fitted angle} & \textbf{Signed $\Delta$} & \textbf{95\% CI of $\Delta$} \\
\midrule
\multirow{2}{*}{2019} & Pleasantness & $340.16^\circ$ & $-19.84^\circ$ & $[-26.03^\circ,-13.52^\circ]$ \\
& Eventfulness & $83.50^\circ$ & $-6.50^\circ$ & $[-10.61^\circ,-2.45^\circ]$ \\
\midrule
\multirow{2}{*}{2025} & Pleasantness & $343.27^\circ$ & $-16.73^\circ$ & $[-23.41^\circ,-9.72^\circ]$ \\
& Eventfulness & $88.24^\circ$ & $-1.76^\circ$ & $[-5.58^\circ,1.95^\circ]$ \\
\bottomrule
\end{tabular}
\end{table}
 
Table~\ref{tab:direct_axis_alignment} reports the fitted direct P and E directions under the two reference geometries. The absolute Pleasantness-axis deviations are $19.84^\circ$ (ISO 2019) and $16.73^\circ$ (ISO 2025), while the corresponding Eventfulness-axis deviations are $6.50^\circ$ and $1.76^\circ$. 
With ISO 2025 as the reference, direct E is fitted at $88.24^\circ$, close to the nominal $90^\circ$ axis, whereas direct P is fitted at $343.27^\circ$, displaced clockwise toward Calm. 
The closer directional correspondence of direct E than direct P is observed under both reference geometries. 
These fitted axes summarise the relationships between direct P/E and the PAQ configuration as a whole, but do not show how individual PAQ attributes are oriented within the direct P/E plane.


\subsection{Direct-anchored PAQ geometry and localised departures}\label{sectionD}

Table~\ref{tab:direct_axis_alignment} describes the fitted direct P and E directions based on all 8 PAQs. To examine the attribute-specific relationships, the direct P/E coordinates define the horizontal ($0^\circ$) and vertical ($90^\circ$) reference axes, allowing each PAQ's direction $\widehat{\theta}_i^{\mathrm{direct}}$ to be estimated from $\beta_{\mathrm{P},i}$ and $\beta_{\mathrm{E},i}$ using Eq.~(\ref{eq:direct_anchored_paq_angle}).

The same 1,782 paired trials are used, with the 8 PAQs participant-centred and standardised by their pooled standard deviations. Unlike the axis analysis in Section~\ref{sectionC}, the 30-s direct P/E means are participant-centred but retain their original common P/E coordinate units. Separate standardisation of P and E would change the angles inferred from their regression coefficients. The coefficient matrix $\mathbf{B}$ is re-estimated for this direct-anchored analysis and is not reused from Section~\ref{sectionC}.

For each PAQ item $i$, the re-estimated coefficients $\beta_{\mathrm{P},i}$ and $\beta_{\mathrm{E},i}$ define a 2D vector in the native direct P/E plane. Its direction is calculated using Eq.~(\ref{eq:direct_anchored_paq_angle}), measured counterclockwise from the positive P axis. For example, Vibrant has coefficients of $0.371$ and $1.420$ for P and E, respectively, giving an estimated direction of $75.37^\circ$. The 8 estimated PAQ directions are then compared with the ISO 2019 and ISO 2025 reference angles.

\begin{figure}[t]
\centering
\includegraphics[width=0.5\columnwidth]{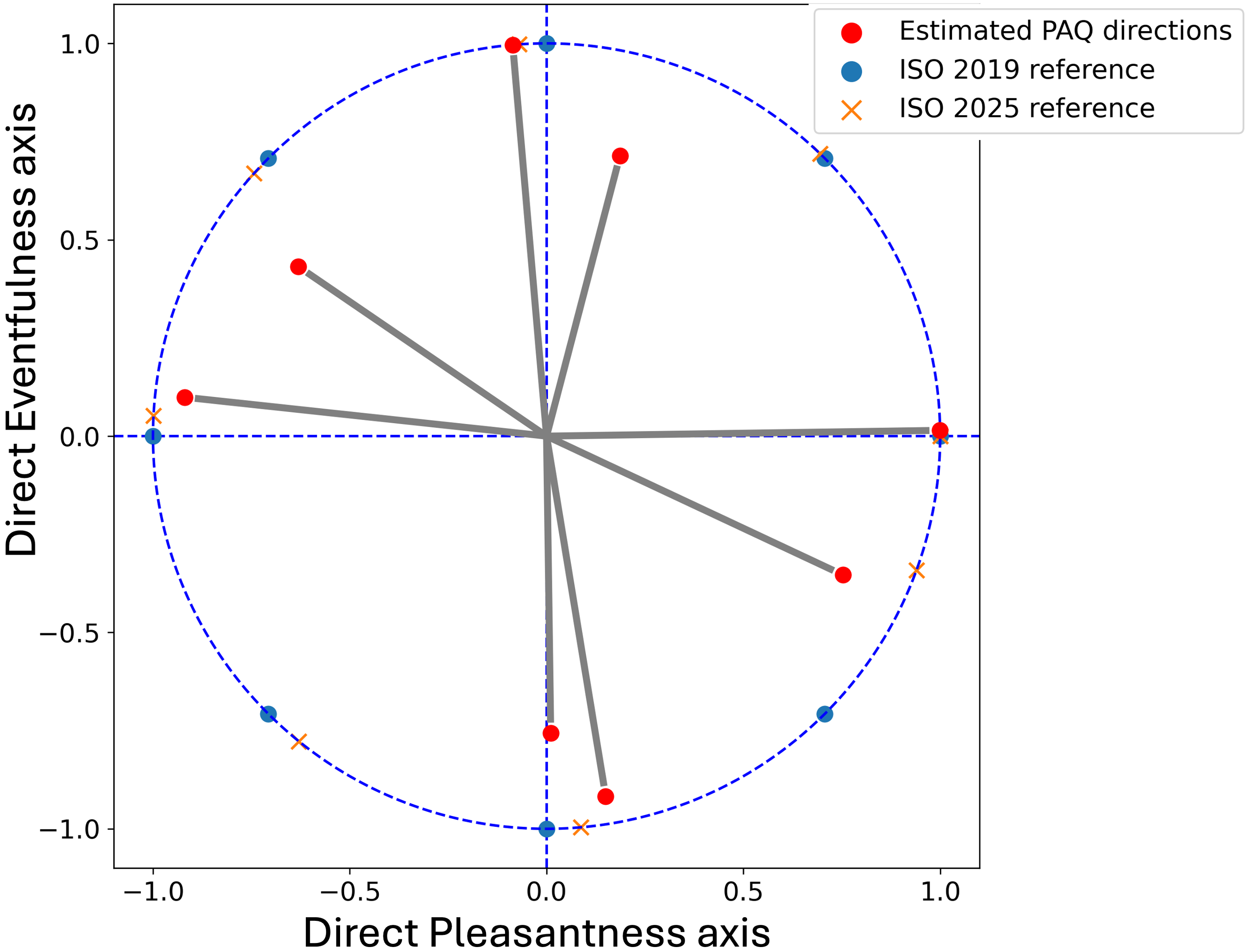}  
\caption{Direct-P/E-anchored geometry of the 8 PAQ attributes. The red endpoint vectors show the estimated PAQ directions, with their lengths scaled by the largest regression-vector magnitude for visualisation. Circles and crosses indicate the ISO 2019 ideal and ISO 2025 adjusted-English reference positions, respectively.}
\label{fig:direct_anchored_paq_geometry}
\end{figure}

Fig.~\ref{fig:direct_anchored_paq_geometry} shows the directions of the 8 retrospective PAQs in the direct P/E plane. Pleasant, Eventful, Chaotic, Annoying, Uneventful, and Calm lie within $7.61^\circ$ of their ISO 2025 reference directions. Vibrant lies at $75.37^\circ$, compared with the reference direction of $46^\circ$, a $+29.37^\circ$ shift toward Eventfulness. Monotonous lies at $270.80^\circ$, compared with $231^\circ$, a $+39.80^\circ$ shift toward Uneventfulness. Both attributes are oriented closer to the Eventfulness--Uneventfulness axis than their ISO 2025 reference directions.

The Vibrant and Monotonous directions are separated by $164.57^\circ$, remaining broadly opposed relative to the expected $180^\circ$. Reversing the Monotonous direction by $180^\circ$ and taking the circular mean of the resulting pair gives a common axis of $83.08^\circ$. The corresponding ISO 2025 reference pair axis is $48.50^\circ$, a difference of $34.58^\circ$. Together with the internal PAQ structure reported in Section~\ref{sec:section_B}, these results answer RQ1: the PAQs retain a predominantly 2D organisation, while Vibrant and Monotonous remain broadly opposed but share an orientation in direct P/E that differs from the ISO 2025 reference geometry.


\subsection{Trial-level agreement between ISO-derived and direct P/E}\label{sectionE}
 
Sections~\ref{sectionC} and~\ref{sectionD} describe the geometric relationships between direct P/E and the retrospective PAQs, including the distinct directions of Vibrant and Monotonous. These geometric relationships do not establish whether the P/E coordinates obtained from the two assessment routes correspond for individual soundscape trials.

\begin{figure}[t]
\centering
\includegraphics[width=0.5\textwidth]{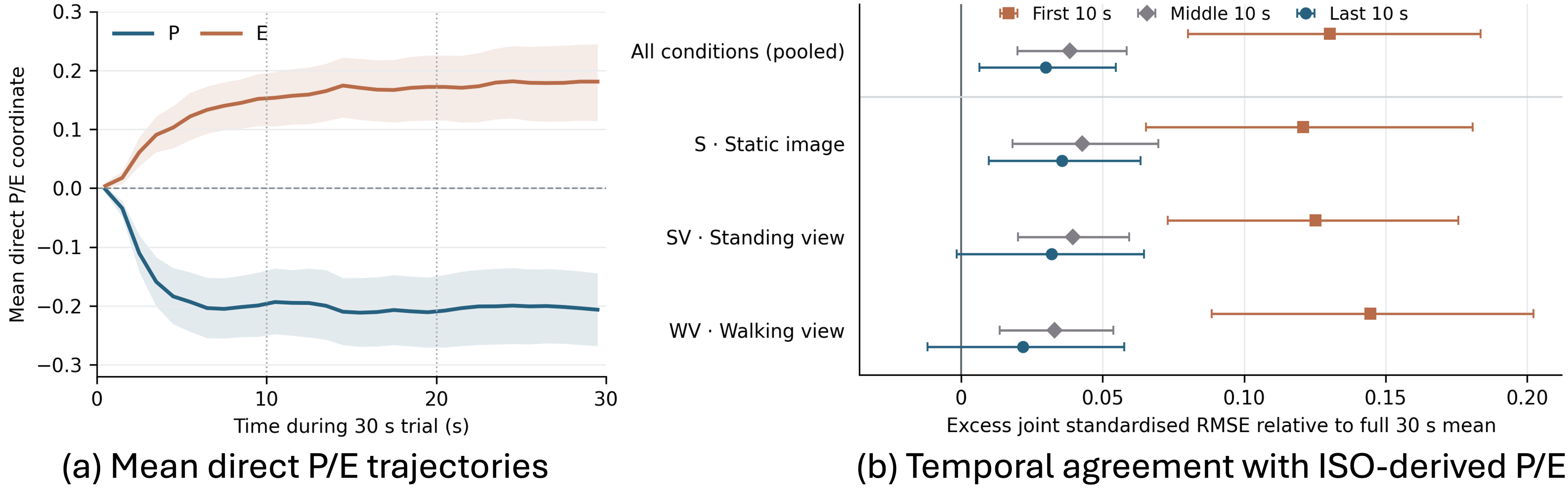}  
\caption{Temporal correspondence between direct and retrospective ISO 2025-derived P/E. Shading and error bars indicate 95\% participant-bootstrap confidence intervals.}
\label{fig:temporal_condition_agreement}
\end{figure}

Each trial provides one ISO-derived P/E coordinate pair from retrospective PAQ ratings, while continuous direct P/E records judgements throughout the 30-s exposure. The correspondence between these measurements may depend on whether the direct P/E trajectory is summarised over the full exposure or a particular portion of it. Four summaries are examined: the full 30-s mean and the first, middle, and last 10-s means. Each is compared with ISO 2025-derived P/E under leave-one-participant-out (LOPO) validation, using a separate affine calibration and joint standardised root mean squared error (RMSE). The temporal analysis uses 1,780 paired trials. Fig.~\ref{fig:temporal_condition_agreement} displays the pooled direct P/E trajectories and compares the temporal summaries overall and for S, SV, and WV.

The pooled direct P/E trajectories in Fig.~\ref{fig:temporal_condition_agreement}(a) change most rapidly during the opening seconds and vary less over the remainder of the exposure. 
Under LOPO validation, joint standardised RMSE is $0.902$ for the full 30-s mean, compared with $1.032$, $0.940$, and $0.932$ for the first, middle, and last 10-s means, respectively. The first 10-s interval has the largest error, while the middle and last 10-s intervals have lower errors but remain above the full-duration mean. Fig.~\ref{fig:temporal_condition_agreement}(b) shows the excess RMSE of the three 10-s summaries relative to the full 30-s mean, with the same point-estimate ordering across S, SV, and WV.
The full-duration means, $\bar{P}_k^{\mathrm{direct}}$ and $\bar{E}_k^{\mathrm{direct}}$ in Eq.~(\ref{eq:direct_mean30}), are used in the main trial-level comparison of all 1,782 paired trials.

\begin{figure}[t]
\centering
\includegraphics[width=0.5\textwidth]{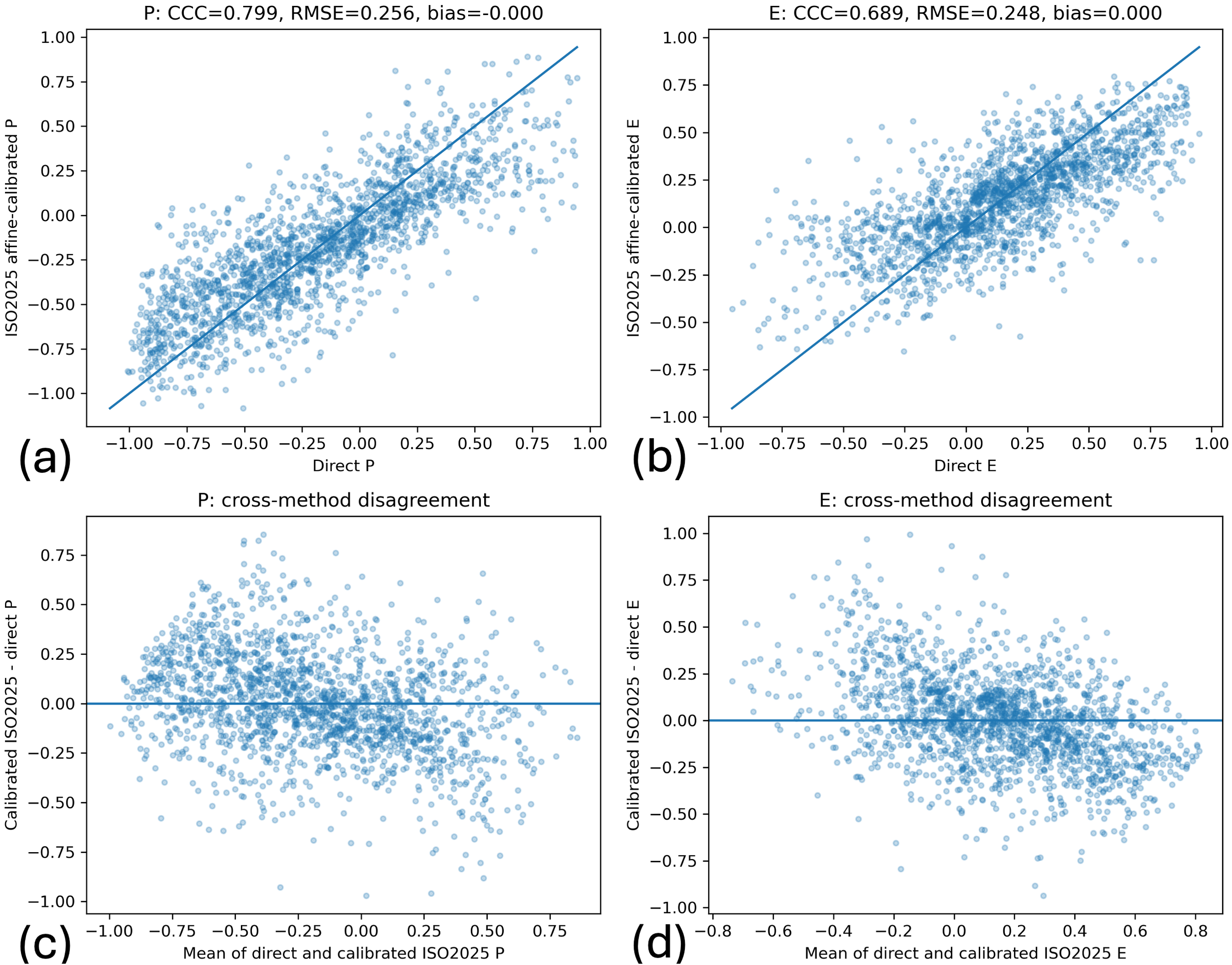}  
\caption{LOPO agreement between calibrated ISO 2025-derived and direct P/E. (a,b) Paired Pleasantness and Eventfulness coordinates with lines of equality. (c,d) Differences between calibrated ISO 2025-derived and direct P/E, calculated as ISO-derived minus direct, plotted against their paired means for Pleasantness and Eventfulness, respectively.}
\label{fig:cross_method_agreement}
\end{figure}

The main trial-level analysis compares the full 30-s direct P/E means with calibrated ISO 2025-derived P/E under both LOPO and LOSO validation. Table~\ref{tab:trial_level_agreement} reports their numerical and positional agreement, while the paired-coordinate and difference-versus-mean plots in Fig.~\ref{fig:cross_method_agreement} show the relationships between the two sets of P/E coordinates across individual trials.

Table~\ref{tab:trial_level_agreement} shows higher concordance for P than E under both validation schemes: LOPO CCC is $0.7986$ for Pleasantness and $0.6885$ for Eventfulness, with a difference of about $0.11$ that persists under LOSO. The coordinate-wise RMSEs plotted in Fig.~\ref{fig:cross_method_agreement}(a,b) are similar for the two dimensions, so comparable error magnitudes do not imply comparable concordance with the line of equality. Joint 2D RMSE is $0.356371$ under LOPO and $0.356420$ under LOSO, showing little change when the held-out unit switches from participants to streets. In the paired-coordinate plots of Fig.~\ref{fig:cross_method_agreement}(a,b), calibrated ISO-derived coordinates occupy a narrower response range than direct P/E, particularly for E. Fig.~\ref{fig:cross_method_agreement}(c,d) shows the same pattern as differences against paired means: calibrated ISO 2025-derived minus direct P/E tends to be positive at lower paired means and negative at higher paired means, with the mean difference near zero after calibration.

\begin{table}[b]
\caption{Trial-level agreement between calibrated ISO 2025-derived and direct P/E.}
\label{tab:trial_level_agreement}
\centering
\small
\setlength{\tabcolsep}{8pt}
\renewcommand{\arraystretch}{1}
\begin{tabular}{@{}lrr@{}}
\toprule
\textbf{Agreement metric} & \textbf{LOPO} & \textbf{LOSO} \\
\midrule
Pleasantness CCC & 0.7986 & 0.7953 \\
Eventfulness CCC & 0.6885 & 0.6880 \\
Joint 2D RMSE & 0.356371 & 0.356420 \\
Quadrant agreement (\%) & 70.37 & 70.54 \\
Median absolute angle ($^\circ$) & 20.5447 & 20.5356 \\
\bottomrule
\end{tabular}
\end{table}

\begin{figure}[t]
\centering
\includegraphics[width=0.5\textwidth]{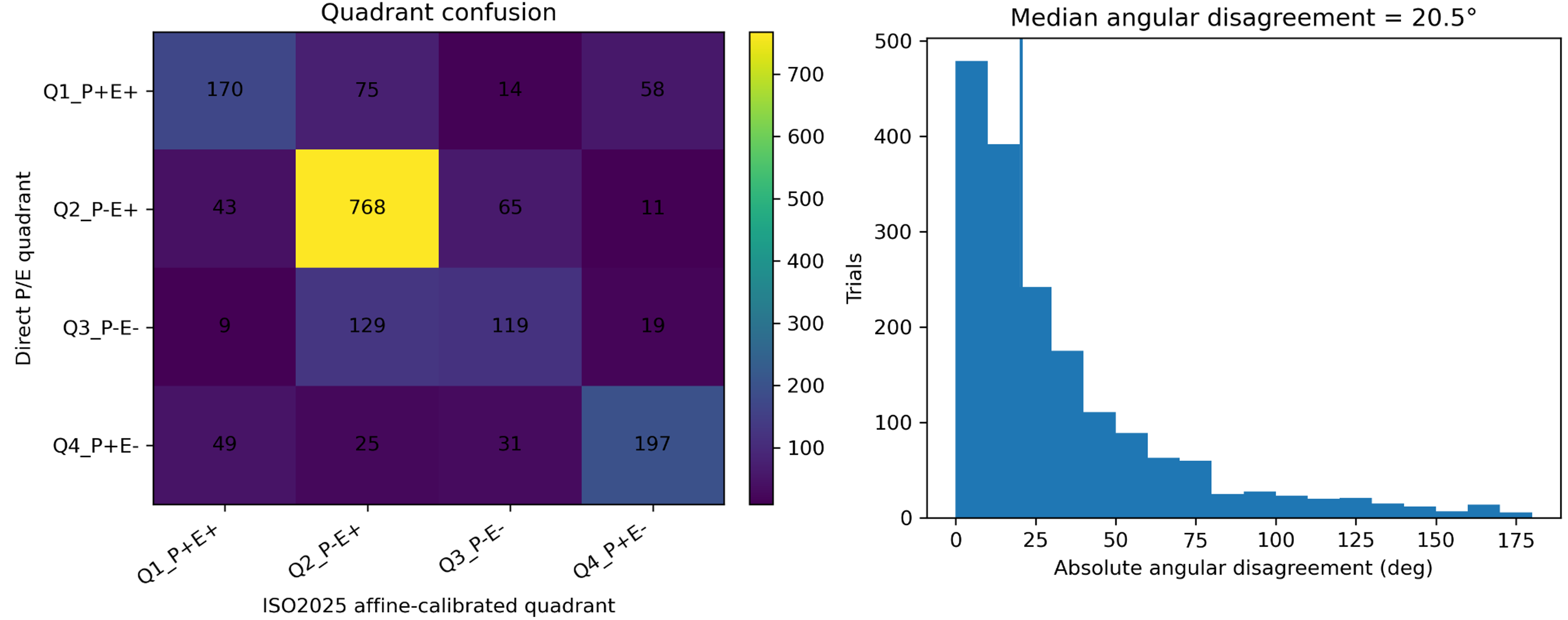}
\caption{LOPO quadrant confusion matrix (left) and distribution of absolute angular differences (right) between calibrated ISO 2025-derived and direct P/E.}
\label{fig:circumplex_consequences}
\end{figure}

The numerical differences also affect the quadrant assigned to an individual trial. Fig.~\ref{fig:circumplex_consequences} shows that calibrated ISO 2025-derived and direct P/E assign 1,254 of the 1,782 paired trials ($70.4\%$) to the same quadrant. The other 528 trials have opposite signs for at least one coordinate. Quadrant assignments can differ even for small coordinate differences near either axis. The confusion matrix shows more sign disagreements for E than for P, consistent with the lower E CCC in Table~\ref{tab:trial_level_agreement}. Under LOPO, the median absolute angular disagreement is $20.54^\circ$, indicating that disagreements in the inferred P/E position extend beyond a simple difference in numerical scale.

These results answer RQ2: the full 30-s direct P/E mean has the lowest LOPO error against retrospective ISO 2025-derived P/E among the four temporal summaries examined, but numerical and quadrant differences remain at the individual-trial level after calibration. In particular, E has lower numerical concordance and more sign disagreements than P, although the independently measured E axis lies close to the nominal ISO 2025 direction in Table~\ref{tab:direct_axis_alignment}.


\subsection{Empirical PAQ-to-P/E mapping from paired assessments}\label{sectionF}
 
Section~\ref{sectionD} identifies differences between individual PAQ directions and their ISO references, particularly for Vibrant and Monotonous, while Section~\ref{sectionE} shows that trial-level differences between ISO 2025-derived and direct P/E remain after affine calibration. The paired measurements also allow an empirical PAQ-to-P/E mapping to be estimated and compared with affine-calibrated ISO 2025 scoring under held-out evaluation. 

The affine-calibrated ISO 2025 coordinate $D_k^{\mathrm{ISO2025-cal}}$ in Eq.~(\ref{eq:iso2025_affine_calibration}) serves as the reference scoring rule. Calibration changes the offset and scale of each coordinate while retaining the ISO-prescribed relative contributions of the 8 PAQs. The relaxed mapping instead estimates the PAQ-to-P/E relationship and the covariance among PAQ responses from paired observations, without fixing the relative PAQ contributions to the ISO 2025 reference geometry.

The relaxed model estimates the relationship through the linear-Gaussian measurement model in Eq.~(\ref{eq:relaxed_measurement_model}). The 30-s direct P/E mean is treated as a noisy observation of the modelled P/E state, not as an error-free target defining that state.

The forward measurement model in Eq.~(\ref{eq:relaxed_measurement_model}) describes retrospective PAQ responses conditional on a modelled P/E state. Its inverse conditional mean in Eq.~(\ref{eq:relaxed_inverse_mapping}) estimates P/E from the 8 retrospective PAQ ratings. Model parameters are fitted using training-fold measurements, and the matched direct P/E observations from held-out trials are reserved for evaluation.

The calibrated ISO 2025 projection and relaxed model are evaluated using identical LOPO and LOSO folds. Table~\ref{tab:relaxed_projection_comparison} reports RMSE separately for P and E and jointly for the 2D coordinates against held-out direct P/E. RMSE is computed within each held-out participant or street and then averaged, giving each fold equal weight. 
This fold-averaged RMSE differs from the trial-pooled RMSE reported in Section~\ref{sectionE}, which combines the held-out trials before calculating RMSE.

\begin{table}[t]
\caption{Fold-averaged held-out RMSE for the ISO 2025 projection and relaxed model.}
\label{tab:relaxed_projection_comparison}
\centering
\small
\setlength{\tabcolsep}{1pt}
\renewcommand{\arraystretch}{0.6}

\begin{tabular}{@{}c|c|c|c|c@{}}
\toprule
\multirow{2}{*}{\textbf{Validation}} &
\multirow{2}{*}{\textbf{Dimension}} &
\textbf{ISO} &
\textbf{Relaxed} &
\multirow{2}{*}{\textbf{Reduction}} \\
& & \textbf{projection} & \textbf{model} & \\
\midrule

\multirow{3}{*}{\textbf{LOPO}}
& P        & 0.2458 & 0.2263 & 7.9\% \\
& E        & 0.2401 & 0.2350 & 2.1\% \\
& Joint 2D & 0.3481 & 0.3311 & 4.9\% \\

\midrule

\multirow{3}{*}{\textbf{LOSO}}
& P        & 0.2554 & 0.2367 & 7.3\% \\
& E        & 0.2455 & 0.2398 & 2.3\% \\
& Joint 2D & 0.3550 & 0.3375 & 4.9\% \\

\bottomrule
\end{tabular}
\end{table}

Table~\ref{tab:relaxed_projection_comparison} shows lower held-out RMSE for the relaxed model under both validation schemes. Under LOPO, joint 2D RMSE decreases from $0.3481$ for affine-calibrated ISO 2025 to $0.3311$ for the relaxed model, a $4.9\%$ reduction. Under LOSO, the corresponding reduction is from $0.3550$ to $0.3375$, also $4.9\%$. The improvement is more pronounced for Pleasantness: under LOPO, P RMSE decreases from $0.2458$ to $0.2263$ ($7.9\%$), whereas E RMSE changes from $0.2401$ to $0.2350$ ($2.1\%$). LOSO retains the same ordering, with reductions of $7.3\%$ for P and $2.3\%$ for E. 

After fitting, the inverse conditional mean in Eq.~(\ref{eq:relaxed_inverse_mapping}) defines an affine mapping from the 8 retrospective PAQ ratings to P/E coordinates without requiring direct P/E recordings for new trials. These results answer RQ3: the empirical mapping estimates the 30-s direct P/E mean with lower held-out RMSE than affine-calibrated ISO 2025 scoring under both LOPO and LOSO validation.

\section{Discussion}\label{sec:discussion}
The relationships among the 8 retrospective perceived affective quality (PAQ) attributes are broadly consistent with the soundscape circumplex, and the fitted direct Pleasantness and Eventfulness (P/E) axes lie near their respective ISO reference directions. However, Vibrant and Monotonous have different directions in the direct P/E plane from those specified by ISO 2025, and the two assessment routes can assign different P/E coordinates to the same exposure. The choice of time interval used to summarise continuous P/E also affects the comparison with retrospective ratings.

\subsection{Perceptual relationships between retrospective PAQs and direct P/E (RQ1)}

The relationships among the 8 retrospective PAQs broadly support the 2D organisation of the soundscape circumplex, consistent with the quasi-circumplex structures observed in listening studies across different languages \cite{aletta2024soundscape}. The first two principal components explain $76.4\%$ of the PAQ variation, and most opposing attribute pairs remain approximately opposite (Fig.~\ref{fig:Fig2_Internal_PAQ_compact}). The internal PAQ directions have lower mean and median absolute deviations from ISO 2019 than from ISO 2025 (Table~\ref{tab:internal_geometry_fit}). These comparisons describe the relationships among retrospective attributes; they do not determine how each attribute relates to P/E reported independently during listening.

The paired direct P/E measurements show how the PAQs relate to Pleasantness and Eventfulness reported during listening. The fitted direct P and E axes lie closer to their nominal directions under ISO 2025 than under ISO 2019 (Table~\ref{tab:direct_axis_alignment}). Relative to ISO 2025, direct E is fitted at $88.24^\circ$, close to $90^\circ$, whereas direct P is fitted at $343.27^\circ$, clockwise from $0^\circ$. Because each fitted axis uses all 8 PAQs, it does not show the direction of any single PAQ in the direct P/E plane.

In the direct-P/E-anchored analysis, six PAQs lie within $7.61^\circ$ of their ISO 2025 reference directions, whereas Vibrant and Monotonous lie closer to the Eventfulness--Uneventfulness axis (Fig.~\ref{fig:direct_anchored_paq_geometry}). Vibrant and Monotonous remain broadly opposed, with a separation of $164.57^\circ$, but their common axis is $83.08^\circ$, compared with $48.50^\circ$ in ISO 2025. The $34.58^\circ$ difference in pair-axis orientation shows that two PAQs can retain an opposing relationship while having different associations with directly reported P/E than their reference directions suggest. 

\subsection{Temporal correspondence and trial-level agreement (RQ2)} 

Comparing coordinates for the same exposure first requires a temporal summary of the continuous direct P/E record. Previous soundscape studies have examined continuous judgements alongside overall ratings, including the contribution of temporal averages and changes during listening to global assessments \cite{aumond2017continuous,graetzer2020continuous}. Here, the retrospective P/E coordinates are calculated from 8 post-exposure PAQ ratings, not from a direct overall P/E judgement. Among the four continuous summaries examined, the full 30-s direct P/E mean gives the lowest calibrated LOPO joint standardised RMSE against ISO 2025-derived P/E ($0.902$), compared with $1.032$, $0.940$, and $0.932$ for the first, middle, and last 10-s means (Fig.~\ref{fig:temporal_condition_agreement}). The last 10-s mean corresponds more closely to retrospective P/E than the first 10-s mean, but neither outperforms the full 30-s mean. The first 10-s interval also contains the largest changes in the pooled direct P/E trajectories (Fig.~\ref{fig:temporal_condition_agreement}), although these observations do not establish how different stages of listening contribute to retrospective appraisal. For comparisons with retrospective PAQ-derived P/E, averaging the full continuous record is better supported by these data than selecting a single 10-s interval. This finding concerns temporal aggregation within a fixed 30-s exposure and does not establish an optimal stimulus duration. Continuous judgements may reflect preceding sounds, and averaging the record removes information about when changes in reported P/E occur. The present analysis does not relate these changes to individual sound events.

Under ISO 2025, the fitted direct E axis lies close to its nominal direction, but the LOPO concordance correlation coefficient between calibrated ISO-derived and direct E is $0.689$, lower than the $0.799$ for P (Tables~\ref{tab:direct_axis_alignment} and~\ref{tab:trial_level_agreement}). The fitted direction describes the association of direct E with the PAQs across trials; concordance describes how closely the two assessment routes assign the same score to each exposure. Calibrated ISO-derived scores also occupy a narrower range than direct P/E, particularly for E, with positive ISO-minus-direct differences at lower paired means and negative differences at higher paired means (Fig.~\ref{fig:cross_method_agreement}).

The remaining numerical differences can change how individual soundscapes are represented in the P/E plane. Calibrated ISO 2025-derived and direct P/E assign $70.4\%$ of trials to the same quadrant (Fig.~\ref{fig:circumplex_consequences}). Because a small coordinate difference near either axis can change the quadrant, the remaining $29.6\%$ of trials do not necessarily involve large differences in P/E scores. When comparing streets or visual conditions by P/E quadrant, the underlying coordinates are needed to distinguish differences near a boundary from larger displacements.

\subsection{Empirical PAQ-to-P/E mapping and score comparability (RQ3)}

The ISO 2025 projection uses prescribed contributions of the 8 PAQs to calculate P/E, and affine calibration adjusts the offset and scale of each resulting coordinate without changing its relative PAQ contributions. The empirical mapping instead estimates the PAQ--P/E relationship and the covariance among PAQ responses using the paired observations. Both approaches take the same 8 retrospective PAQ ratings as input and are evaluated against held-out direct P/E. The 30-s direct P/E means are treated as observations with measurement variation, not as error-free perceptual states.

Under both LOPO and leave-one-street-out (LOSO) validation, the empirical mapping reduces fold-averaged held-out joint RMSE by $4.9\%$ relative to affine-calibrated ISO 2025 (Table~\ref{tab:relaxed_projection_comparison}). Under LOPO, the reduction is larger for P ($7.9\%$) than for E ($2.1\%$), although E has lower trial-level concordance in the comparison of the two assessment routes. The held-out results do not isolate the contributions of individual PAQs or covariance terms to the improvement, and numerical differences from direct P/E remain.

After fitting, the empirical mapping estimates a P/E coordinate pair from the 8 retrospective PAQ ratings without requiring continuous direct P/E for a new trial. It does not reconstruct the direct P/E trajectory, and the LOPO and LOSO results concern held-out participants and streets within the same experiment. The ISO projection provides a common scoring rule for retrospective PAQ assessments, whereas an empirical mapping characterises their relationship with direct P/E in the data used to estimate it. Whether that relationship is consistent across independently collected soundscapes requires separate evidence. For comparisons across soundscapes, the common P/E plane does not guarantee the same scores under different assessment procedures. 

\section{Conclusion}\label{sec:Conclusion}  
Soundscape research uses the 8 perceived affective quality (PAQ) attributes as a common language for describing perceived acoustic quality. ISO-derived Pleasantness and Eventfulness (P/E) coordinates provide a standardised representation of retrospective PAQ ratings, while continuous direct P/E recorded during listening provides an independent, time-resolved description of the same soundscape exposure.

Across the 1,782 paired 30-s trials, the PAQs retain a predominantly 2D arrangement, and the direct P/E axes broadly follow the ISO reference geometry. In the direct P/E plane, Vibrant and Monotonous lie closer to the Eventfulness--Uneventfulness axis than their ISO 2025 reference directions. These geometric relationships coexist with numerical differences between the two assessment routes for individual exposures. In the temporal analysis of 1,780 paired trials, the full 30-s mean of continuous direct P/E has lower calibrated LOPO joint standardised RMSE against ISO 2025-derived P/E than any of the three 10-s summaries. Under LOPO validation with affine calibration, the two routes assign 29.6\% of the 1,782 paired trials to different P/E quadrants. An empirical PAQ-to-P/E mapping fitted to the paired observations reduces fold-averaged held-out joint RMSE by 4.9\% relative to affine-calibrated ISO 2025 under both LOPO and LOSO validation, using the same 8 retrospective PAQ ratings as inputs. The improvement is modest, and differences from direct P/E remain.

Future work could test whether the temporal relationships observed in 30-s exposures hold for other durations, whether the numerical correspondence between the two assessment routes varies across visual conditions, and how individual sound events relate to retrospective ratings. The two assessment routes also differ in rating time and response format; adding post-exposure direct P/E ratings could help distinguish these effects. Independent paired datasets are needed to test whether the observed numerical differences persist and whether the empirical PAQ-to-P/E mapping generalises to other soundscapes and listeners. Within this experiment, retrospective PAQs and continuous direct P/E provide complementary descriptions of the same exposures, and their numerical correspondence cannot be inferred from PAQ geometry alone.





\bibliography{JASA}

\end{document}